\documentclass[a4paper,UKenglish,cleveref, autoref, thm-restate]{lipics-v2021}

\pdfoutput=1 
\hideLIPIcs  
\nolinenumbers

\title{Generalizing Weighted Path Orders}

\author{Teppei Saito}
{JAIST, Japan}
{saito@jaist.ac.jp}
{https://orcid.org/0009-0001-9786-0044}
{JST SPRING, Grant Number JPMJSP2102}

\author{Nao Hirokawa}
{JAIST, Japan}
{hirokawa@jaist.ac.jp}
{https://orcid.org/0000-0002-8499-0501}
{JSPS KAKENHI Grant Number JP22K11900}

\authorrunning{T. Saito and N. Hirokawa}
\Copyright{Teppei Saito and Nao Hirokawa}


\category{} 

\relatedversion{} 

\acknowledgements{We are grateful to Vincent van Oostrom for his valuable
questions and comments on our preliminary work.  We also thank Alfons Geser
for his support on literature.  The comments of the anonymous reviewers 
helped to improve the presentation of the paper.}

\EventEditors{John Q. Open and Joan R. Access}
\EventNoEds{2}
\EventLongTitle{42nd Conference on Very Important Topics (CVIT 2016)}
\EventShortTitle{CVIT 2016}
\EventAcronym{CVIT}
\EventYear{2016}
\EventDate{December 24--27, 2016}
\EventLocation{Little Whinging, United Kingdom}
\EventLogo{}
\SeriesVolume{42}
\ArticleNo{23}

\usepackage{multicol}
\usepackage{stmaryrd}
\usepackage{ebproof}
\usepackage{xspace}
\usepackage{booktabs}

\newcommand{\m}[1]{\mathsf{#1}}

\newcommand{\seq}[2][n]{{#2_1},\dots,{#2_{#1}}}

\newcommand{\bN}{\mathbb{N}}
\newcommand{\bZ}{\mathbb{Z}}
\newcommand{\cA}{\mathcal{A}}

\newcommand{\cF}{\mathcal{F}}

\newcommand{\cR}{\mathcal{R}}

\newcommand{\cT}{\mathcal{T}}
\newcommand{\cV}{\mathcal{V}}

\newcommand{\lexstar}{\mathsf{lex}} 

\newcommand{\spo}{\mathsf{spo}}
\newcommand{\mspo}{\mathsf{mspo}}
\newcommand{\wpo}{\mathsf{wpo}}
\newcommand{\gwpo}{\mathsf{gwpo}}

\newcommand{\WPO}[1]{\textrm{WPO\,#1}\xspace}
\newcommand{\SPO}[1]{\textrm{SPO\,#1}\xspace}

\newcommand{\sqsupsetsim}{\mathrel{\vphantom{\gtrsim}\smash{\ooalign{\raise.4ex\hbox{$\sqsupset$}\cr$\raise-.85ex\hbox{$\sim$}$}}}}

\begin{document}

\maketitle

\begin{abstract}
We show that weighted path orders are special instances of a variant of
semantic path orders.  Exploiting this fact, we introduce a
generalization of weighted path orders that goes beyond the realm of simple
termination.  Experimental data show that generalized weighted
path orders are viable.
\end{abstract}

\section{Introduction}
\label{sec:introduction}
Various classes of reduction orders have been proposed since reduction orders
are a fundamental tool in termination analysis of term rewrite systems and
in completion-based 
theorem proving.  Concurrently, several
attempts have been made to unify these classes of reduction orders.  Among
others, \emph{weighted path orders} (WPOs)~\cite{YKS15} are
known as a powerful generalization of Knuth--Bendix orders (KBOs).
Extending notion of weight function to \emph{simple monotone
algebra}, WPOs unify KBOs and lexicographic path orders (LPOs).

Another approach toward unification is to generalize parameters of so-called path orders
so as to take orders on terms, such as
semantic path orders (SPOs)~\cite{KL80},
\emph{monotonic semantic path orders} (MSPOs)~\cite{BFR00,B03}, and
general path orders (GPOs).
However, relationship between WPOs and these approaches
has not been thoroughly investigated,
except for the one between extended KBOs and SPOs~\cite{G92}.
In particular, the relationship between WPOs and MSPOs has remained open~\cite{Y18}.

In this note, we demonstrate an effective construction of an MSPO
from the algebra and the precedence of a given WPO.
This simulation result leads to a generalization of WPOs
that does not impose simplicity on their underlying algebras.
The fact that the generalization can show termination of term rewrite systems
that are not simply terminating
forms a sharp contrast from the original ones.
Besides, WPOs have been implemented in several tools for termination
analysis and automated theorem proving.  Upgrading WPOs to GWPOs can be
done with little implementation effort, so we anticipate that these tools
may benefit from power of GWPOs.

The rest is organized as follows:
In \cref{sec:wpo} we recall weighted path orders.
In \cref{sec:spo} it is shown that weighted path orders can be simulated by
a variant of semantic path orders.
\cref{sec:gwpo} is devoted to a generalization of weighted path orders and
discussion on its experimental data.
\cref{sec:future-work} concludes the note with future work.

This note is a short version of our recent conference paper~\cite{SH23}.

\section{Weighted Path Orders}
\label{sec:wpo}
First we recall notions related to weighted path orders, see also the
original work~\cite{YKS15}.
Terms are constructed from the finite signature $\cF$ and variables
$\cV$, and the set of terms is denoted by $\cT(\cF, \cV)$.
We may write $f^{(n)}$ for a function symbol $f$ in order to indicate that
$f$ has the arity $n$.
WPOs comprise two ingredients.
One is a \emph{precedence},
which is a quasi-order on the signature.
The other
is an \emph{ordered algebra} $\cA = ({A}, {\{f_\cA\}_{f \in \cF}}, >)$.
Here $>$ is a strict order on a set $A$,
and $f_\cA$ is an $n$-ary function on $A$ associated with each $f^{(n)} \in \cF$.
We write $s >_\cA t$ if $[\alpha]_\cA(s) > [\alpha]_\cA(t)$ for all assignments $\alpha$.
Similarly, we write $s \geqslant_\cA t$
if $[\alpha]_\cA(s) \geqslant [\alpha]_\cA(t)$ for all assignments $\alpha$
where $\geqslant$ is the reflexive closure of $>$.
The ordered algebra $\cA$ is
\begin{itemize}
\item
\emph{simple} if
$f_\cA(a_1, \ldots, a_i, \ldots, a_n) \geqslant a_i$ for all
$f^{(n)} \in \cF$, $1 \leqslant i \leqslant n$, and
$\seq{a} \in A$;
\item
\emph{weakly monotone} if
\(
f_\cA(a_1,\ldots,a_i,\ldots,a_n) \geqslant
f_\cA(a_1,\ldots,b,\ldots,a_n)
\)
for all 
$f^{(n)} \in \cF$, argument positions $1 \leqslant i \leqslant n$, and
$\seq{a}, b \in A$ with $a_i > b$;
\item
\emph{simple monotone} if it is simple and weakly monotone; and
\item
\emph{well-founded} if $>$ is well-founded.
\end{itemize}

\begin{definition}[\textnormal{\cite{YKS15}}]
\label{def:wpo}
Let $\cA$ be an ordered $\cF$-algebra and $\succsim$ a precedence.
The \emph{weighted path order} $>_\wpo$ is defined
on terms as follows: $s >_\wpo t$ if
\begin{enumerate}
\item  $s >_\cA t$, or
\item $s \geqslant_\cA t$, $s = f(\seq[m]{s})$, and one of the following
conditions holds.
\begin{enumerate}[a.]
\item
$s_i \geqslant_\wpo t$ for some $1 \leqslant i \leqslant m$.
\item \smallskip
$t = g(\seq{t})$ and $s >_\wpo t_j$ for all $1 \leqslant j \leqslant n$, and
moreover
\smallskip
    \begin{enumerate}[(i)]
    \item
    $f \succ g$, or
    \item
    $f \succsim g$ and $(\seq[m]{s}) >_\wpo^\lexstar (\seq{t})$.
    \end{enumerate}
\end{enumerate}
\end{enumerate}
Here $\geqslant_\wpo$ denotes the reflexive closure of $>_\wpo$.
\end{definition}

\begin{theorem}[\textnormal{\cite{YKS15}}]
\label{thm:wpo}
For every simple monotone well-founded algebra and precedence
the induced relation $>_\wpo$ is a reduction order.
\end{theorem}

A TRS $\cR$ is terminating if and only if there exists a reduction order $>$
with $\cR \subseteq {>}$.  Therefore, WPOs can be used for showing
termination of TRSs.

\begin{example}
\label{ex:wpo}
Consider the following TRS
\(
    \cR = \{
    \m{f}(\m{g}(x)) \to \m{g}(\m{f}(\m{f}(x))),
    \m{f}(\m{h}(x))  \to \m{h}(\m{h}(\m{f}(x)))
    \}
\) taken from~\cite[Example~9]{YKS15}.
Let $\cA$ be the simple monotone algebra on $\bN$ with
$\m{f}_\cA(x) = \m{h}_\cA(x) = x$ and $\m{g}_\cA(x) = x + 1$.
Take a precedence $\succsim$ with $\m{f} \succ \m{g} \succ \m{h}$.
The relation
$\m{f}(\m{g}(x)) >_\wpo \m{g}(\m{f}(\m{f}(x)))$ is verified
by the following derivation:
\begin{center}
\begin{prooftree}
    \hypo{\m{f}(\m{g}(x)) \geqslant_\cA \m{g}(\m{f}(\m{f}(x)))}
    \hypo{\m{f} \succ \m{g}}
    \hypo{\m{f}(\m{g}(x)) >_\cA \m{f}(\m{f}(x))}
    \infer{1}[\WPO{1}]{\m{f}(\m{g}(x)) >_\wpo \m{f}(\m{f}(x))}
\infer{3}[\WPO{2b(i)}]{\m{f}(\m{g}(x)) >_\wpo \m{g}(\m{f}(\m{f}(x)))}
\end{prooftree}
\end{center}
Here \WPO{1} and \WPO{2b(i)} indicate the corresponding conditions in
\Cref{def:wpo}.
Besides, one can verify
$\m{f}(\m{h}(x)) >_\wpo \m{h}(\m{h}(\m{f}(x)))$.
Hence, we conclude that $\cR$ is terminating.
\end{example}

In the case that the signature is finite, a rewrite order $>$ is
called a \emph{simplification order} if it has
the \emph{subterm property},
that is, $s > t$ holds whenever $t$ is a proper subterm of a term $s$.
For instance, WPOs satisfying the conditions of \cref{thm:wpo} are simplification orders.
A TRS $\cR$ is \emph{simply terminating} if
it has a simplification order $>$ with $\cR \subseteq {>}$.
The termination proving power of WPOs as a reduction order is
characterized by simple termination.
Needless to say, not every terminating TRS is simply terminating.
Such instances are considered in \cref{sec:gwpo}.

\section{Simulating WPOs by SPOs}
\label{sec:spo}

Borralleras~\cite[Definition~4.1.19]{B03} introduced
a variant of SPO that employs a pair of a quasi-order and a strict order.
This variant compares arguments of terms by a multiset order.  In order to
simulate WPOs which compare arguments in a lexicographic manner, we
introduce another variant of SPO.

We say that the pair $({\gtrsim},{>})$ of a quasi-order $\gtrsim$ and a
strict order $>$
is an \emph{order pair} if
$\gtrsim \cdot > \cdot \gtrsim {\subseteq} >$.
We say that an order pair $({\sqsupsetsim},{\sqsupset})$ on $\cT(\cF, \cV) \setminus \cV$
is \emph{stable} if both $\sqsupsetsim$ and $\sqsupset$ are stable.

\begin{definition}
\label{def:spo}
Let $({\sqsupsetsim}, {\sqsupset})$ be a stable order pair on $\cT(\cF, \cV) \setminus \cV$.
The \emph{semantic path order  $>_\spo$ (SPO)} is defined on terms as follows:
$s >_\spo t$ if $s = f(\seq[m]{s})$ and one of the following conditions hold:
\begin{enumerate}
    \item
    $s_i \geqslant_\spo t$ for some $1 \leqslant i \leqslant m$. 
    \item
    $t = g(\seq{t})$ and $s >_\spo t_j$ for all $1 \leqslant j \leqslant n$,
    and moreover
    \begin{enumerate}[a.]
    \item
    $s \sqsupset t$, or
    \item
    $s \sqsupsetsim t$ and $(\seq[m]{s}) >_\spo^{\lexstar} (\seq{t})$.
    \end{enumerate}
\end{enumerate}
Here $\geqslant_\spo$ denotes the reflexive closure of $>_\spo$.
\end{definition}

In general, semantic path orders are not closed under contexts.  For a
remedy, Borralleras et al.~\cite{BFR00} propose the use of another preorder
with the \emph{harmony} property.
This results in monotonic semantic path orders.

\begin{definition}[\textnormal{\cite[Definition~4.1.19]{B03}}]
A triple $({\gtrsim},{\sqsupsetsim}, {\sqsupset})$  is a \emph{reduction triple} if
$\gtrsim$ is a rewrite preorder on terms,
$({\sqsupsetsim},{\sqsupset})$ is a stable order pair on $\cT(\cF, \cV) \setminus \cV$
with $\sqsupset$ well-founded, and
$\gtrsim$ and $\sqsupsetsim$ have the \emph{harmony} property, meaning that for every 
$f^{(n)} \in \cF$ the implication 
\(
s_i \gtrsim t \implies
f(s_1,\ldots,s_i,\ldots,s_n) \sqsupsetsim f(s_1,\ldots,t,\ldots,s_n)
\)
holds for all terms $\seq{s},t$ and argument positions
$1 \leqslant i \leqslant n$.
\end{definition}

\begin{definition}
Let $({\gtrsim},{\sqsupsetsim}, {\sqsupset})$ be a reduction triple,
and let $>_\spo$ be the semantic path order induced from $({\sqsupsetsim}, {\sqsupset})$.
The \emph{monotonic semantic path order $s >_\mspo t$ (MSPO)} is defined as
$s \gtrsim t$ and $s >_\spo t$.
\end{definition}

\begin{theorem}
\label{thm:mspo}
Every monotonic semantic path order is a reduction order. \qed
\end{theorem}

We construct  a suitable order pair $({\sqsupsetsim},{\sqsupset})$
from the weakly monotone well-founded
algebra $\cA$ and the precedence $\succsim$ of a WPO $>_\wpo$. 
For terms $s = f(\seq[m]{s}), t = g(\seq{t})$ we write $s \sqsupsetsim t$ if
$s >_\cA t$, or both $s \geqslant_\cA t$ and $f \succsim g$.  Similarly, we
write $s \sqsupset t$ if $s >_\cA t$, or both $s \geqslant_\cA t$ and 
$f \succ g$.

\begin{lemma}
The pair $({\sqsupsetsim},{\sqsupset})$ is a stable order pair with
$\sqsupset$ well-founded.
\qed
\end{lemma}

\begin{lemma}
\label{lem:quasi-reduction-triple}
The triple $({\geqslant_\cA}, {\sqsupsetsim},{\sqsupset})$ forms a quasi-reduction triple.
\qed
\end{lemma}

It is worth noting that $\sqsupset$ is not always
the strict part of $\sqsupsetsim$.
This is why the variant of SPOs is introduced.

\begin{example}
Let the signature be $\{ \m{f}^{(1)} \}$. 
Consider the trivial
precedence $\m{f} \succsim \m{f}$ and the algebra $\cA$ over the carrier
$\bN$ with the interpretation $\m{f}_\cA(x) = 2 x$.  On the one hand we have
$\m{f}(\m{f}(x)) \sqsupsetsim \m{f}(x)$ from
$\m{f}(\m{f}(x)) \geqslant_\cA \m{f}(x)$
but not $\m{f}(x) \sqsupsetsim \m{f}(\m{f}(x))$ as 
$\m{f}(x) \ngeqslant_\cA \m{f}(\m{f}(x))$.
On the other hand $\m{f}(\m{f}(x)) \sqsupset \m{f}(x)$ does not hold.
So $\sqsupset$ is not the strict part of $\sqsupsetsim$.
\end{example}

Let $>_\spo$ be the SPO induced from $({\sqsupsetsim},{\sqsupset})$
and $>_\mspo$ the MSPO induced from $({\geqslant_\cA}, {\sqsupsetsim},{\sqsupset})$.
The following theorem is shown by a straightforward induction proof.

\begin{theorem}
\label{thm:WPO-MSPO}
The three orders $>_\wpo, >_\spo, >_\mspo$ coincide, provided that
$\cA$ is simple. 
\qed
\end{theorem}

\begin{example}[continued from \Cref{ex:wpo}]
\label{ex:spo}
From $\m{f}(\m{g}(x)) \geqslant_\cA \m{g}(\m{f}(\m{f}(x)))$ and 
$\m{f} \succ \m{g}$ the orientation
$\m{f}(\m{g}(x)) \sqsupset \m{g}(\m{f}(\m{f}(x)))$ is obtained.
Moreover, we have $\m{f}(\m{g}(x)) >_\cA \m{f}(\m{f}(x))$.  Because
$\geqslant_\cA$ has the subterm property, the subterm $\m{f}(x)$ of
$\m{f}(\m{f}(x))$ also satisfies $\m{f}(\m{g}(x)) >_\cA \m{f}(x)$.
So we obtain $\m{f}(\m{g}(x)) \sqsupset \m{f}(\m{f}(x)), \m{f}(x)$.  Therefore,
$\m{f}(\m{g}(x)) >_\spo \m{g}(\m{f}(\m{f}(x)))$ is verified as follows:
\begin{center}
\begin{prooftree}
    \hypo{\makebox[7em][l]{$\m{f}(\m{g}(x)) \sqsupset \m{g}(\m{f}(\m{f}(x)))$}}
    \hypo{\makebox[6em][l]{$\m{f}(\m{g}(x)) \sqsupset \m{f}(\m{f}(x))$}}
    \hypo{\m{f}(\m{g}(x)) \sqsupset \m{f}(x)}
        \hypo{x \geqslant_\spo x}
    \infer{1}[\SPO{1}]{\m{g}(x) \geqslant_\spo x}
    \infer{1}[\SPO{1}]{\m{f}(\m{g}(x)) >_\spo x}
    \infer{2}[\SPO{2a}]{\m{f}(\m{g}(x)) >_\spo \m{f}(x)}
    \infer{2}[\SPO{2a}]{\m{f}(\m{g}(x)) >_\spo \m{f}(\m{f}(x))}
\infer{2}[\SPO{2a}]{\m{f}(\m{g}(x)) >_\spo \m{g}(\m{f}(\m{f}(x)))}
\end{prooftree}
\end{center}
In addition, $\m{f}(\m{h}(x)) >_\spo \m{h}(\m{h}(\m{f}(x)))$ can be verified.
Hence, the inclusion $\cR \subseteq {>_\spo}$ holds.
Observe that the use of \WPO{1} in \cref{ex:wpo} is replaced by \SPO{1} and \SPO{2a}.
\end{example}

\section{Generalized Weighted Path Orders}
\label{sec:gwpo}
\Cref{thm:WPO-MSPO} states that weighted path orders can be 
defined as monotonic semantic path orders. Moreover,
\Cref{lem:quasi-reduction-triple} reveals that even
for non-simple algebras the construction of reduction triples is
valid.  This suggests a generalization of weighted path orders,
which does not impose simplicity on algebras.  Besides, we exploit the fact
that stable order pairs need not be closed under contexts, marking root
symbols of function applications~\cite[Definition~5]{BFR00}.

Let $\cF$ be a signature.  For each $f \in \cF$ we associate a marked
function symbol $f^\sharp \notin \cF$ of the same arity.
The set $\{ f^{\sharp} \mid f \in \cF \}$ is denoted by $\cF^\sharp$.
For each term $t = f(\seq{t})$ in $\cT(\cF,\cV) \setminus \cV$
we denote $f^\sharp(\seq{t})$ by $t^\sharp$.
Let $\cA$ be a weakly monotone well-founded $(\cF \cup \cF^\sharp)$-algebra 
and $\succsim$ a precedence on $\cF$.
The pair $({\sqsupsetsim}, {\sqsupset})$ of relations on $\cT(\cF,  \cV) \setminus \cV$
is defined as follows: Let $s = f(\seq{s}), t = g(\seq[m]{t})$.
We write $s \sqsupsetsim t$ if $s^\sharp >_{\cA} t^\sharp$,
or $s^\sharp \geqslant_{\cA} t^\sharp$ and $f \succsim g$;
Similarly we write $s \sqsupset t$ if $s^\sharp >_{\cA} t^\sharp$, or $s^\sharp \geqslant_{\cA} t^\sharp$ and $f \succ g$.
The relation $\gtrsim$ is defined as the restriction of $\geqslant_\cA$ to $\cT(\cF, \cV)$.

\begin{proposition}
The triple $({\gtrsim}, {\sqsupsetsim}, {\sqsupset})$ is a reduction triple on $\cT(\cF, \cV)$.
\qed
\end{proposition}

\begin{definition}
\label{def:gwpo}
The \emph{generalized weighted path order (GWPO)} $>_\gwpo$ induced from
$\cA$ and $\succsim$ is the monotonic semantic path order induced from
$({\gtrsim}, {\sqsupsetsim}, {\sqsupset})$.
\end{definition}

\begin{corollary}
\label{cor:gwpo}
Every generalized weighted path order is a reduction order.
\qed
\end{corollary}

\begin{corollary}
The relations $>_\gwpo$ and $>_\wpo$ coincide,
provided that $\cA$ is simple and $f_\cA(\seq{x}) = f^\sharp_\cA(\seq{x})$ for all $f^{(n)} \in \cF$.
\qed
\end{corollary}

The following two examples illustrate termination proofs by
\cref{cor:gwpo}.  Neither of the TRSs is simply terminating,
and thus their termination cannot be shown by WPOs.

\begin{example}
\label{ex:div}
Consider the TRS $\cR$ for round-up division:
\begin{align*}
\m{p}(\m{0}) & \to \m{0}
&
x - \m{0} & \to x
&
\m{0} \div \m{s}(y) & \to \m{0}
\\
\m{p}(\m{s}(x)) & \to x
&
x - \m{s}(y) & \to \m{p}(x) - y
&
\m{s}(x) \div \m{s}(y) & \to \m{s}((x - y) \div \m{s}(y))
\end{align*}
Let $\cA$ be the weakly monotone algebra on $\bN$ with the
interpretations
\begin{align*}
\m{0}_\cA & = 0
& \m{s}_A(x) & = x + 1
& \m{p}_\cA(x) & = x
& x -_\cA y & = x
& x \div_\cA y & = x
\\
\m{0}^\sharp_\cA & = 0
& \m{s}^\sharp_\cA(x) & = 0
& \m{p}^\sharp_\cA(x) & = 0
& x -^\sharp_\cA y & = y
& x \div^\sharp_\cA y & = x + y
\end{align*}
and let $\succsim$ be an arbitrary precedence.  The GWPO induced from $\cA$
and $\succsim$ orients all rules in $\cR$. Hence, $\cR$ is terminating.
\end{example}

\begin{example}
\label{ex:bits}
Consider the TRS $\cR$ (\texttt{Strategy\_removed\_AG01\_\#4.28} from
TPDB~11.3):
\begin{xalignat*}{3}
\m{half}(\m{0}) & \to \m{0}
&
\m{half}(\m{s}(\m{0})) & \to \m{0}
&
\m{half}(\m{s}(\m{s}(x))) & \to \m{s}(\m{half}(x))
\\
\m{bits}(\m{0}) & \to \m{0}
&
\m{bits}(\m{s}(x)) & \to \m{s}(\m{bits}(\m{half}(\m{s}(x))))
\end{xalignat*}
Let $\cA$ be the weakly monotone algebra on $\bN$
with:
\begin{align*}
\m{0}_\cA & = 0
&
\m{s}_\cA(x) & = x + 1
&
\m{half}_\cA(x) & = \max \{ 0, x - 1 \}
&
\m{bits}_\cA(x) & = x
\\
\m{0}^\sharp_\cA & = 0
&
\m{s}^\sharp_\cA(x) & = x + 1
&
\m{half}^\sharp_\cA(x) & = \max \{ 0, x - 1 \}
&
\m{bits}^\sharp_\cA(x) & = x
\end{align*}
The GWPO $>_\gwpo$ induced by $\cA$ and a precedence $\succsim$ with $\m{half}, \m{bits} \succ \m{s}$ satisfies
$\cR \subseteq {>_\gwpo}$. So $\cR$ is terminating.
\end{example}

In order to evaluate GWPOs in termination analysis we implemented a
prototype termination tool based on \cref{cor:gwpo}.
Following the automation techniques of WPO~\cite{YKS15}, we search a
suitable weakly monotone well-founded algebra from two classes of
algebras over $\bN$.
One is \emph{linear interpretation} and the other is
\emph{max/plus interpretation}.
Since simplicity of algebras is not required for GWPOs, we may use
more general forms of interpretations.

Algebras $\cA$ of linear interpretations use linear polynomials over $\bN$ like \cref{ex:div}.
For each $f^{(n)} \in \cF \cup \cF^\sharp$ its interpretation is of the form
$f_\cA(x_1, \ldots, x_n) = c_0 + c_1 x_1 + \cdots + c_n x_n$
where $c_0 \in \bN$
and $c_1, \ldots, c_n \in \{ 0, 1 \}$.\footnote{%
If we allow $c_0 < 0$ by extending the carrier to $\bZ$,
the well-foundedness of GWPOs is lost.
In fact, when $\m{f}_\cA(x) = \m{f}^\sharp_\cA(x) = x-1$ and
$\m{g}_\cA(x) = \m{g}^\sharp_\cA(x) = x-2$, we have
$\m{f}(x) >_\gwpo \m{f}(\m{g}(x)) >_\gwpo \m{f}(\m{g}(\m{g}(x))) >_\gwpo \cdots$.}
Simple monotone algebras for WPOs are obtained by setting
$c_1 = \cdots = c_n = 1, f_\cA = f^{\sharp}_\cA$ for all $f^{(n)} \in \cF$,
and those for Knuth--Bendix orders (KBOs) are 
obtained by further restriction for admissibility, see \cite{YKS15}.

Algebras $\cA$ of max/plus interpretations use a combination of $+$ and $\max$ like \cref{ex:bits}.
For each $f^{(n)} \in \cF \cup \cF^\sharp$ its interpretation is of the form
$f_\cA(x_1, \ldots, x_n) = \max\{c_0, c_1 + d_1 x_1, \cdots, c_n + d_n x_n\}$
where $c_0 \in \bN, c_1, \ldots, c_n \in \bZ$ and $d_1, \ldots, d_n \in \{ 0, 1 \}$.
Simple monotone algebras for WPOs are obtained by 
imposing
$c_1, \ldots, c_n \in \bN, d_1 = \cdots = d_n = 1, f_\cA = f^\sharp_\cA$
for all $f^{(n)} \in \cF$, and
algebras for lexicographic path orders (LPOs) are obtained
by further restriction $c_0 = c_1 = \cdots = c_n = 0$ for all $f^{(n)} \in \cF$ as in \cite{YKS15}. 
The restriction $c_1, \ldots, c_n \in \bN$ is necessary for WPOs
because allowing $c_1, \ldots, c_n \in \bZ$ results in non-simple interpretations
such as $\max \{0, x -1\}$.

The problem set consists of 1511 term rewrite systems 
from version 11.3 of the Termination Problem Database (TPDB).
The reference implementation uses the SMT solver Z3 as an
external tool for solving linear constraints.
The experiments were run on a PC with Intel Core i7-1065G7 CPU
(1.30 GHz) and 16 GB memory. 

Now let us discuss the experimental results.\footnote{%
The implementation and the detailed experimental data are available at:
\url{https://www.jaist.ac.jp/project/maxcomp/23frocos/}}
\cref{tbl:experiments} shows
that,
as a whole, use of non-simple algebras substantially improves termination analysis,
at the small cost of extra running time.
In particular, in the case of linear interpretation, 
GWPOs significantly outperform WPOs. As a matter of fact, 
linear WPOs are unable to orient variable duplicating rules $\ell \to r$ such as $\m{f}(x) \to \m{g}(x, x)$
since $\ell \geqslant_\cA r$ cannot be satisfied,
but this does not apply to GWPOs based on linear interpretations
which allow $0$ as coefficients.
In the case of max/plus interpretations
there are two TRSs (with over 100 rules) that are proved to be terminating
by WPOs, but not by GWPOs due to the time limit.
This indicates that using non-simple algebras for max/plus interpretation
can result in increase of search space.  This is not the case for linear
interpretations.

\begin{table}[t]
\caption{Experiments on 1511 TRSs from TPDB 11.3.}
\label{tbl:experiments}
\centering
\begin{tabular}{l@{\hspace{3em}}r@{\quad}r@{\quad}r@{\hspace{3em}}r@{\quad}r@{\quad}r}
\toprule
interpretations
& \multicolumn{3}{c@{\qquad}}{\textit{linear}}
& \multicolumn{3}{c@{\qquad}}{\textit{max/plus}}
\\
order
& KBO
& WPO
& GWPO
& LPO
& WPO
& GWPO
\\
\midrule
proved TRSs
& 103
& 122
& 357
& 149
& 221
& 385
\\
\textit{timeouts} (60 sec)
& \textit{8}
& \textit{9}
& \textit{9}
& \textit{12}
& \textit{12}
& \textit{28}
\\
\bottomrule
\end{tabular}
\end{table}

\section{Future Work}
\label{sec:future-work}

We have introduced a generalization of WPOs whose termination proving power
goes beyond the realm of simple termination. We conclude the paper by
discussing future work.

\subparagraph*{General path orders.}
In this paper only the lexicographic versions of path orders were
investigated.  However, it is very likely that the same result can be
obtained even if we adopt multiset comparison or status functions.  General
path orders (GPOs) are a unifying framework for such extensions,
parameterizing the way to compare arguments.  It is worth investigating
simulation results between GPOs and WPOs by extending the parameters of
GPOs so as to take order pairs.

\subparagraph*{Reduction pairs based on WPOs.}
In order to build reduction pairs for Arts and Giesl's dependency pair
method, Yamada et al.~\cite[Section~4]{YKS15} extended the definition of
WPOs by the notion of \emph{partial status function}.  
The resulting reduction-pair version of WPOs does not impose the
simplicity condition on algebras.  It is not difficult to see that whenever
termination of a TRS is shown by a GWPO then it can also be shown by the
dependency pair method with a reduction pair based on WPOs.  We
anticipate that the converse also holds if we integrate the notion of
partial status functions into GWPOs.



\bibliography{references}

\end{document}